\documentclass[10pt, twocolumn]{article}

\usepackage{graphicx}
\usepackage{caption}
\usepackage{subcaption}
\usepackage{gensymb}
\usepackage{float}
\usepackage{wrapfig}
\usepackage{gensymb}
\usepackage{amsmath}
\usepackage{amssymb}
\usepackage{rotating}
\usepackage{enumitem}
\usepackage{bm}
\usepackage{multicol}
\usepackage{color}
\usepackage{cite}
\usepackage{url}
\usepackage{hyperref}
\usepackage{setspace}
\usepackage{lastpage}
\usepackage[top=3.35cm, bottom=3.35cm, left=2.25cm, right=2.25cm]{geometry}
\usepackage{fancyhdr}
\usepackage{sectsty}
\definecolor{Grey}{rgb}{0.4, 0.4, 0.4}
\definecolor{Red}{rgb}{1.0, 0.0, 0.0}
\sectionfont{\normalsize\bfseries}
\subsectionfont{\normalsize\itshape}
\subsubsectionfont{\normalsize\itshape}
\begin{document}
\begin{singlespace}
\twocolumn[\begin{center}
\begin{large}
\textbf{Development of a High Fidelity Simulator for Generalised Photometric Based Space Object Classification using Machine Learning}\\
\end{large}
\vspace{0.5cm} 
\textbf{James Allworth\textsuperscript{1,2}, Lloyd Windrim\textsuperscript{1}, Jeffrey Wardman\textsuperscript{2,3}, Daniel Kucharski\textsuperscript{3}, James Bennett\textsuperscript{2,3},  Mitch Bryson\textsuperscript{1}}
\\
\bigskip
\textbf{1} \textit{Australian Centre for Field Robotics, The University of Sydney, Sydney, Australia, \\Email: \{j.allworth, l.windrim, m.bryson\}@acfr.usyd.edu.au}
\\
\textbf{2} \textit{EOS Space Systems Pty Ltd, Canberra, Australia, Email: \{jallworth, jwardman, jbennett\}@eosspacesystems.com}
\\
\textbf{3} \textit{Space Environment Research Centre, Canberra, Australia, Email: \{jeffreywardman, jamesbennett, danielkucharski\}@serc.org.au}
\\
\bigskip
\bigskip

\textbf{Abstract}
\vspace{-0.2cm}
\end{center}
This paper presents the initial stages in the development of a deep learning classifier for generalised Resident Space Object (RSO) characterisation that combines high-fidelity simulated light curves with transfer learning to improve the performance of object characterisation models that are trained on real data. The classification and characterisation of RSOs is a significant goal in Space Situational Awareness (SSA) in order to improve the accuracy of orbital predictions. The specific focus of this paper is the development of a high-fidelity simulation environment for generating realistic light curves. The simulator takes in a textured geometric model of an RSO as well as the objects ephemeris and uses Blender to generate photo-realistic images of the RSO that are then processed to extract the light curve. Simulated light curves have been compared with real light curves extracted from telescope imagery to provide validation for the simulation environment.  Future work will involve further validation and the use of the simulator to generate a dataset of realistic light curves for the purpose of training neural networks.
\bigskip
\bigskip]

\textbf{Acronyms}\\
RSO - Resident Space Object\\
SSA - Space Situational Awareness\\
EOS - Electro Optic Systems\\
TLE - Two Line Element\\
SGP4 - Simplified General Perturbations 4\\
TEME - True Equator, Mean Equinox\\
SERC - Space Environment Research Centre\\
GUI - Graphical User Interface\\
FFT - Fast Fourier Transform

\section{Introduction}
\vspace{-0.2cm}
Space Situational Awareness (SSA) is becoming increasingly important as a rise in the accessibility of space couples with the difficulty in removing space debris objects.  Due to the high relative velocites between objects in orbit, even small pieces of debris ($< 5$cm) pose a significant threat to operational spacecraft.  The problem of space debris is exacerbated by a collisional cascading effect, where collisions between satellites create debris increasing the chance of more collisions \cite{kessler_collision_1978}.\\

In order to mitigate the risk that space debris poses to operational satellites, various organisations maintain catalogues of on-orbit objects. Observations of Resident Space Objects (RSOs) are used in an orbit determination process to estimate their orbital state, which can then be propagated to predict possible conjunctions with other RSOs within a specific time period.  This information is valuable to satellite operators if it is accurate, precise and timely.\\  

Orbital perturbations caused by non-conservative forces, such as drag and solar radiation pressure, can reduce the accuracy of these predictions and are dependent on an object's size, shape, attitude and reflectivity properties \cite{dianetti_observability_2018}.  If these parameters can be determined (via RSO characterisation) they can be used to increase the accuracy of the RSO orbital prediction. Additionally, RSO characterisation is useful in identifying potential targets for debris removal campaigns, with rapidly rotating objects more likely to experience break up events creating more debris.\\

This paper presents the initial stages in the development of a deep learning classifier for generalised RSO characterisation using ground-based light curve measurements.
A light curve is a series of discrete measurements of the object's brightness over time \cite{silha_apparent_2018}, in intensity or magnitude units, which can be extracted from optical observations of an RSO.  A light curve can be acquired for any object that is illuminated by the Sun and is reflecting enough light in the direction of a ground based observer to be visible for instruments.  The extraction and analysis of a number of light curves for different objects \cite{silha_apparent_2018, linder_extraction_2015, fruh_analysis_2010},  has demonstrated that for rotating objects, a repeating pattern can often be observed in the light curve.\\

Previous research has shown that light curves are a function of an RSO's physical characteristics, including its size, shape, orientation and material properties, as well as the distance to the observer and the phase angle relative to the sun and observer \cite{jah_satellite_2007, wetterer_attitude_2009, linares_inactive_2012, linares_space_2014}. Traditionally, the methods for recovering the RSOs characteristics from light curve data have been heavily reliant on estimation theory, which have a solid theoretical background but tend to be computationally expensive \cite{furfaro_space_2018}.\\ 

Since the early 2010s, breakthroughs made in the deep learning community, largely as a result of more powerful computers, larger datasets and improved techniques to train deeper networks, have led to remarkable results on a number of perceptual problems \cite{goodfellow_deep_2017}.  Deep learning is a subset of machine learning. A deep neural network typically comprises of several layers of non-linear function approximators that are trained to learn representations of data useful for tasks such as classification \cite{chollet_deep_2018}.  The recent widespread success of deep learning solutions to a variety of problems has motivated their application across numerous fields including RSO characterisation\cite{linares_space_2016, furfaro_space_2018, jia_space_2018, huo_classification_2019}.\\

In 2016, Linares and Furfaro \cite{linares_space_2016} argued that the traditional approaches to RSO characterisation involved a large compuational burden which would not be practical for the ever increasing catalogue of RSOs.  They theorised that a data-driven deep learning approach would enable quick determination of RSO classes made directly from observational data.  While results on simulated datasets have shown success \cite{furfaro_space_2018, linares_space_2016}, application to real light curve data has proven to be more difficult due to the smaller dataset size and the differences between the simulated and real light curve data \cite{furfaro_space_2018}.\\

Existing studies in other fields such as computer vision have shown that deep learning models are able to outperform traditional methods when training sets are large \cite{krizhevsky_imagenet_2017}.  However observational RSO light curve data is both difficult to obtain and label as a number of the objects have unknown characteristics.  Consequently, the size and quality of the training dataset limits the achievable performance and robustness of the neural network.\\

A high-fidelity light curve simulation environment enables the possibility of creating a large, well-labelled dataset that could be used to supplement observational RSO light curve datasets for the purposes of training neural networks.  Provided the simulated light curves encapsulate similar features to those found in a real light curve dataset, models pre-trained on the simulated dataset could then be fine-tuned on a small real light curve dataset in a process known as transfer learning.  Transfer learning is known to be an effective way of increasing the performance of neural networks, particularly in cases where the dataset is small \cite{mahajan_exploring_2018}.\\

Previous research applying neural networks to simulated light curves and investigating transfer learning \cite{furfaro_space_2018, linares_space_2016, furfaro_resident_2016} has used a simplified version of the anisotropic-Phong bidirection reflectance distribution function (BRDF) developed by Ashihman and Shirley \cite{ashikhmin_anisotropic_2000} to model reflectance.  This has then been applied to simple spacecraft and debris models represented through a combination of flat facets to simulate light curves.  Using this flat facet method it is difficult to create complex models that provide realistic representations of RSOs.  Additionally, this method does not incorporate self shadowing or secondary reflections and the use of flat facets creates edges in curved surfaces. \\ 

By creating a more realistic simulation environment using recent advancements in rendering and computer graphics, it is hypothesised that the simulated light curves will encapsulate more of the features observed in real light curve data and hence result in the development of more accurate and robust RSO characterisation models based on transfer learning. The specific contributions of our paper are:
\begin{itemize}
\item Development of Blender based simulation environment to generate a realistic light curve dataset that can be used for transfer learning.
\item Validation of the simulation environment through comparison with real data. 
\end{itemize}

Section 2 of this paper presents an overview of the simulation environment as well as providing details about its three main elements. Section 3 outlines the algorithm developed to automatically extract real light curves from Electro Optic Systems (EOS) optical data. The results of the light curve extraction algorithm are presented in Section 4 and are compared with simulated light curves to validate the simulation environment.  Section 5 provides a discussion of the work presented and Section 6 contains a brief conclusion of the research presented as well as ideas for future work.

\section{Methodology}
\begin{figure*}
	\centering
	\includegraphics[width=\textwidth]{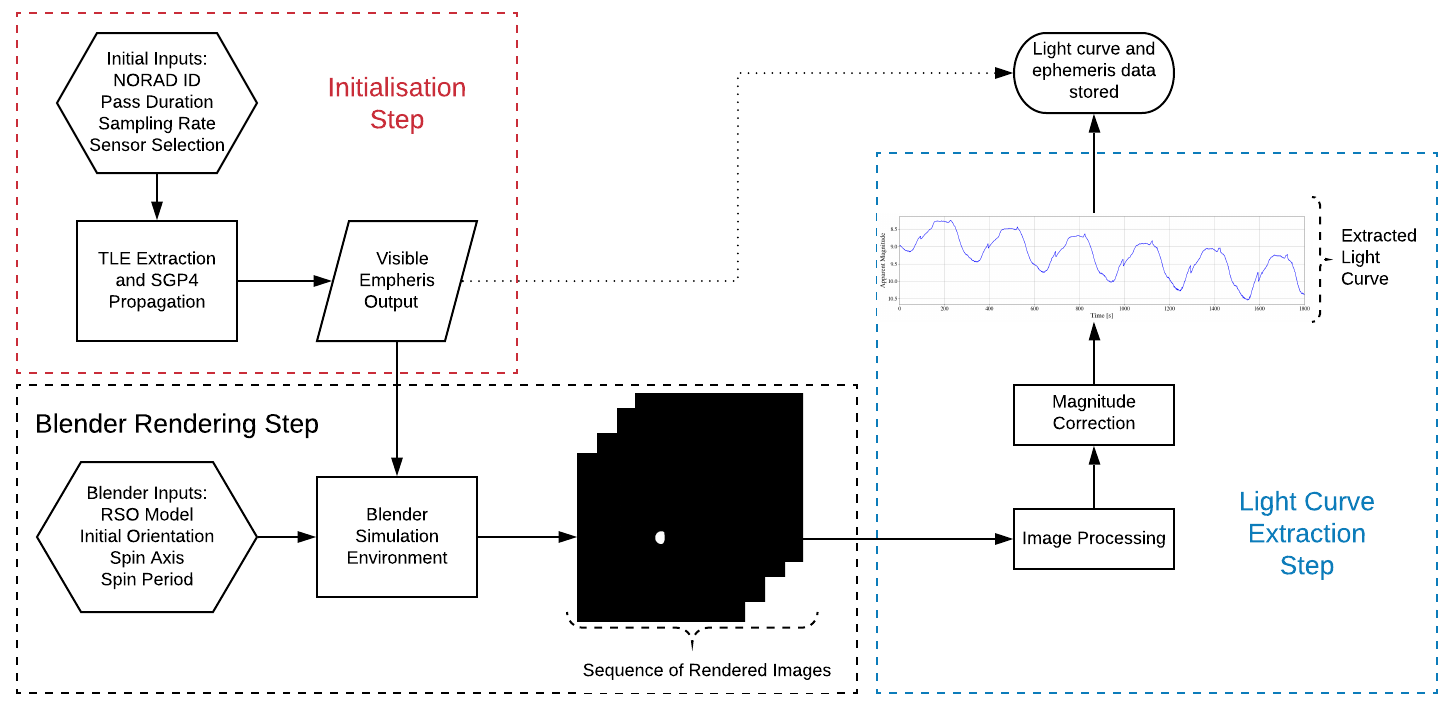}
	\caption{Simulation Environment Pipeline}
	\label{fig:sim_pipeline}
\end{figure*}

\vspace{-0.2cm}
A light curve simulator has been developed that takes a textured geometric model of a given RSO, the RSO orbital parameters and ground-based measuring location, and generates the simulated measurement of a light curve (apparent magnitude vs. time), based on the solar reflection from the object. The simulator pipeline is depicted in Figure \ref{fig:sim_pipeline} and is composed of three main steps:
\begin{enumerate}
\item Initialisation Step: Uses RSO ephemeris to determine when it is visible from a specified sensor and records position information for the RSO, sensor and sun at a specified sampling rate.
\item Blender Rendering Step: Uses the ephemeris information recorded in the previous step to render photo-realistic images of the solar illuminated RSO from the perspective of the ground-based observer. 
\item Light Curve Extraction Step: Processes the rendered images to produce simulated measurements of single-point apparent magnitude from rendered data.
\end{enumerate}

\subsection{Initialisation Step}
\vspace{-0.2cm}
The aim of the simulation environment is to accurately simulate the real world geometry of the three primary components, the RSO, the ground-based sensor and the sun, in order increase the fidelity of the simulation and to reduce the difficulty in comparing real and simulated light curves.  Subsequently, the Blender environment requires the input of position information for these three components to be recorded at each time step.\\

This initialisation step is performed in Python where a graphical user interface (GUI) allows the user to select the NORAD ID of the object as well as the start epoch, pass duration, sampling rate and ground-based sensor. The Two Line Element (TLE) ephemeris for the object is automatically extracted from the Space-Track website with the closest TLE prior to the start of the simulated pass selected.  This TLE is then propagated using the Simplified General Perturbations 4 (SGP4) propagator in conjunction with the Naval Observatory Vector Astrometry Software (NOVAS) library to obtain the position information of the sun, the observer and the RSO at each time step. To simplify coordinate transformations, the origin of the Blender coordinate system was considered to be the centre of mass of the earth which is the origin of the true equator, mean equinox (TEME) coordinate system.  TEME was selected as it is the coordinate system used by TLEs and SGP4.\\

At each time step, a visibility check is performed to ensure that the object is illuminated by the sun and visible to the observer. The conditions to determine visibility for an observer are as follows:
\begin{itemize}
\item RSO elevation $> 15\degree$ above horizon from the perspective of the observing sensor. At lower elevations the reflected light travels through more of the atmosphere making it difficult to acquire the RSO and obtain accurate measurements.
\item Sun elevation $< -6\degree$ below horizon from the perspective of the observing sensor.  For the optical observations to be recorded the observing sensor should not be illuminated by the sun.  
\item RSO not in earth shadow.  The RSO must be illuminated by the sun for it to be visible to the observer. This was calculated through implementation of the conic shadow model provided by Hubaux \cite{hubaux_symplectic_2012}.
\end{itemize}

When the object is determined to be visible to the observer the vectorised position information of the sun, RSO and sensor are recorded in the TEME coordinate system.  If the RSO is not visible no information is recorded and it is propagated forward in time until it becomes visible or reaches the end of the pass specified by the user. The recorded ephemeris information is used as input to initialise the next stage in the simulation pipeline.

\subsection{Blender Rendering Step}
\vspace{-0.2cm}
Previous research has shown Blender to be an effective tool for simulating realistic light curves through comparisons between real data and Blender simulated light curves\cite{kanzlar_space_2015, linder_extraction_2015}.  Blender is an open source 3D rendering software, which has a built-in physically based probabilistic ray-tracing engine called Cycles \cite{blender}.  At the time of writing Blender 2.80 was the most updated stable release, so it was the version used for the following simulations. \\

The simulation environment consists of three primary components in Blender; the sun, the camera and the object models.  The position information for these components is input into Blender using the ephemeris calculated in the previous step.  Due to issues determined in Blender when rendering dim objects at extreme distances, a scale reduction of 1000 was applied equally to the position information for all three components.  This scale reduction is corrected for in the light curve extraction step so that an accurate comparison can be made with real world data.\\

Additional inputs controlling the selection of the object model as well as the initial orientation and rotation of the object model are also required at this point.   This can either be input in manual mode, which allows a user to control the inputs for simulating a specific pass or in simulation mode where these parameters are randomly selected from a range of possible options.

\subsubsection{Sun}
\vspace{-0.2cm}
To simulate the sun a default Blender sun lamp was used, which provides light of a constant intensity emitted in a single direction from infinitely far away.  Consideration was included for the colour of the sunlight by setting the black body shader node in Blender to a temperature of 5778K, which controls the colour of the light from the sun node.  The intensity of the sunlight impacting the model was set to be 1062W/m$^2$.  This was determined to be the intensity of the sunlight in the visible spectrum impacting on objects in earth orbit \cite{mehta_photometric_2018}.  No consideration is made for small changes in intensity at different altitudes and all incoming light rays are considered parallel.

\subsubsection{Camera}
\vspace{-0.2cm}
Blender provides a default camera object with a number of options for customisation.  This allowed multiple cameras to be created with the aim of simulating the various sensors that EOS used to collect the real data.  These camera objects are then saved into the Blender file and can be selected during the simulation process. All cameras are set to record imagery in grey-scale as is the case for EOS sensors.  Additionally, the OpenEXR format was used to output the simulated imagery with the gamma correction set to 1.0 and a linear transform used to prevent the corrections that Blender normally applies to image data for visual effect.  The cameras were also simulated to have the same position as the EOS sensors at each epoch to facilitate ease of comparison with real data.\\  

\subsubsection{Object Models}
\vspace{-0.2cm}
The  GUI in Blender allows the modelling and visualisation of complex objects and is developed to handle highly detailed models with a range of materials and textures.  In a similar manner to the cameras, object models stored in the Blender file format can be selected to be used during the simulation.  The materials used for the models are also controllable during the simulation runtime as opposed to having to store the same object multiple times to test the effect of different materials on the light curve.\\

\begin{figure}
	\centering
	\includegraphics[width=1.0\columnwidth]{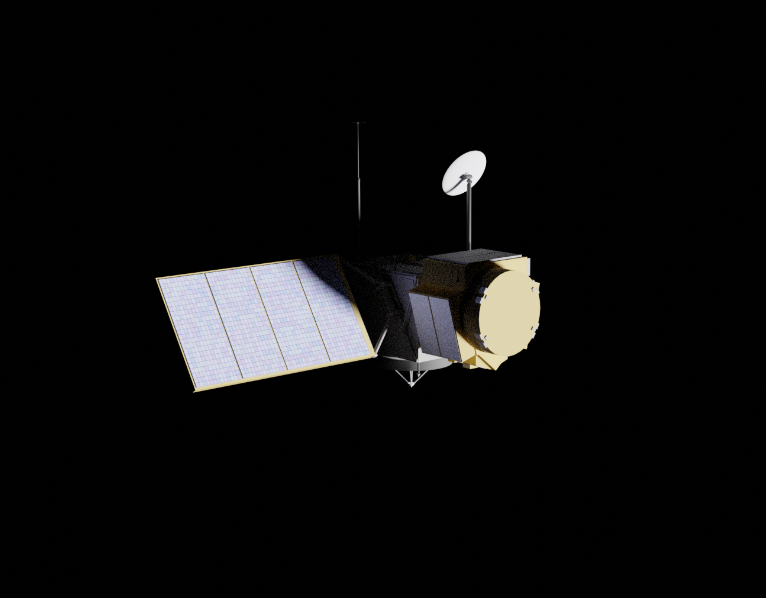}
	\caption{Topex Model}
	\label{fig:Topex}
\end{figure}

Additionally, Blender supports a wide range of common 3D formats, allowing object models developed in other programs to be imported into the simulation environment.  This has been useful for initial simulation testing as a number of models have been downloaded from the NASA 3D model website \cite{nasa_models}.  Figure \ref{fig:Topex} depicts a rendering of the Topex model downloaded from the NASA website as an example of the kind of models used during the simulation.\\

Figure \ref{fig:Topex} also demonstrates that self shadowing is incorporated into Blender as well as secondary reflections where light is reflected from one surface onto another. It is important to incorporate self shadowing for models containing concave or extruding features such as solar panels.

\subsubsection{Motion Blur Implementation}
\vspace{-0.2cm}
During initial testing of the simulation it was determined that to emulate real light curves the object must continue to rotate during the image exposure as it does in the real world. Previous simulations using the flat facet method rotate the object to a certain orientation and use the normal vectors of the facets at this orientation to calculate the light reflected in the direction of the observer.  However, for objects that are rapidly rotating or for simulations involving long exposures, the change in orientation of the normal vectors during exposure affects the intensity of the light reflected towards the observer.\\

\begin{figure}
	\centering
	\begin{subfigure}{0.49\columnwidth}
		\includegraphics[width=\linewidth]{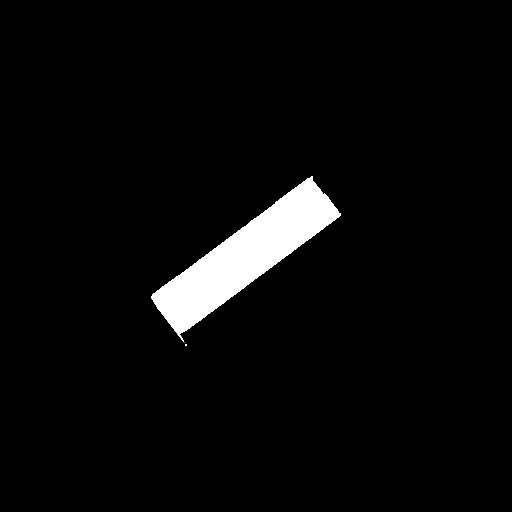}
		\caption{No Motion Blur}
		\label{fig:still}
	\end{subfigure}
	\begin{subfigure}{0.49\columnwidth}
		\includegraphics[width=\linewidth]{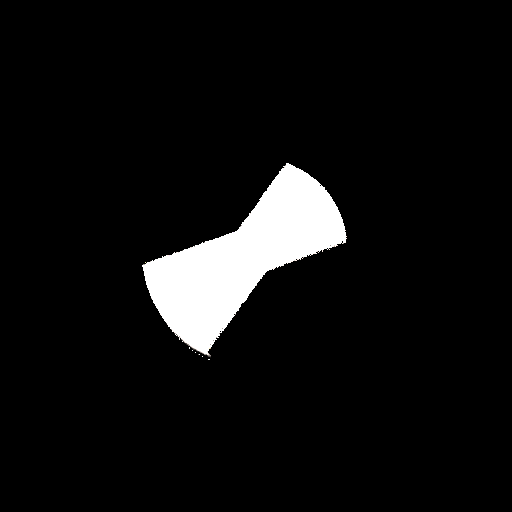}
		\caption{Motion Blur}
		\label{fig:blur}
	\end{subfigure}
	\caption{Example of Motion Blur Implementation}
	\label{fig:blur_image}
\end{figure}

This effect can be simulated in Blender by applying animation to the object and motion blur to the scene.  Figure \ref{fig:blur_image} depicts a comparison between the rendering of an object with and without motion blur.  The simulated object is a 3U CubeSat rotating around its x axis with  spin period of 10 seconds, scaled to be resolvable in the image. Further analysis on the impact that this implementation has on the extracted light curve is presented in the results section.

\subsection{Light Curve Extraction Step}
\vspace{-0.2cm}
After the full pass has been rendered into a sequence of images, the images are input into the light curve extraction step to determine the apparent magnitude of the object in each image. Image processing is used to determine the brightness of the object in the image through integration over the pixel counts. However converting this Blender intensity of the object to a physically accurate apparent magnitude is non-trivial and requires some understanding or the apparent magnitude system.\\

The apparent magnitude is a measure of the relative brightness of an object as seen by an observer on earth based on an inverse logarithmic relation defined by the following equation. 
\begin{equation}
m = -2.5\; \textrm{log}_{10}(f)
\end{equation} 
Where $m$ is the apparent magnitude and $f$ is the photon flux of the object received at the observer.\\

Consequently, very bright objects have lower magnitudes, such as the sun with a magnitude of -26.7, whilst dimmer objects have higher magnitudes.  Significant work has been performed in the field of astronomy to build star catalogues with various information about each star recorded including their apparent magnitudes.  The apparent magnitude of a star can be determined by comparing the flux of that star with the flux of another star with a known magnitude as follows \cite{romanishin_introduction_2006}:
\begin{equation}
m_1 = m_2 - 2.5\; \textrm{log}_{10} \left(\frac{f_1}{f_2} \right)
\label{eq:mag_comp}
\end{equation}

It is difficult to determine the exact photon flux in an instrument due to calibration requirements.  However, the instrument magnitude for an object can be recorded using the pixel counts in an image as outlined in \cite{romanishin_introduction_2006}:

\begin{equation}
I_m = -2.5\; \textrm{log}_{10} \left(\frac{N_{ap} - A_{ap}S_{sky}}{t_{exp}} \right)
\label{eq:inst_mag}
\end{equation}

Where $N_{ap}$ is the sum of the pixels in the aperture, $A_{ap}$ the area of the aperture (in pixels), $S_{sky}$ is the background sky signal and $t_{exp}$ is the exposure time of the image (in seconds).\\

By replacing the flux values in Equation \ref{eq:mag_comp} with the relationship inside the log in Equation \ref{eq:inst_mag} and then applying log laws the resultant expression can be derived to determine the apparent magnitude of an object from the pixel values in an image through comparison with another object with a known apparent magnitude.
\begin{equation}
m_1 = m_2 + I_{m_1} - I_{m_2} 
\label{eq:final_rel}
\end{equation}
Where $m_1$ is the apparent magnitude of the object being determined, $m_2$ is the known apparent magnitude of an object (usually a star) and  $I_{m_1}$ and $I_{m_2}$ are the calculated instrument magnitudes of the two objects in the image.\\

Thus in the real world the apparent magnitude of an RSO is determined by comparing its instrument magnitude with that of a star with a known apparent magnitude through the applications of Equations \ref{eq:inst_mag} and \ref{eq:final_rel}.  However, this is not possible in a simulation because there is no zero point star in the system with a known magnitude to compare other objects against.\\

In the case of the flat facet model, this is not an issue because no image is actually produced. Instead the model takes the intensity of the sunlight as an input in $W/m^2$ and calculates the fraction of this light that would reach the observer after correction for distance.  See \cite{furfaro_space_2018} for further detail on this method where the authors state that the apparent magnitude of an object in the simulation can be calculated through application of Equation \ref{eq:mag_comp} as follows:
\begin{equation}
m = -26.7 - 2.5\; \textrm{log}_{10} \left( \frac{F}{C_{sun,vis}} \right)
\label{eq:mag_fur}
\end{equation}
Where -26.7 is the value for the apparent magnitude of the sun, $F$ is the calculated fraction of the reflected sunlight from the object that is visible to the observer and $C_{sun,vis}$ is the intensity of the sunlight in earth orbit.\\

Subsequently, the flat facet model was used to create a reference point for converting the extracted sum of pixel intensities from the Blender-generated images into a corrected apparent magnitude that could be compared to real data.  A simple 1$\times$1\,m flat plate with the same material settings, was simulated at a distance of 10000\,m in both Blender and the flat facet model using the same object, sun and camera geometry to ensure that reflective angles were the same.\\

Using the method outlined in \cite{furfaro_space_2018} and through application of Equation \ref{eq:mag_fur}, the apparent magnitude of the flat plate from the flat facet method was determined.  The instrument magnitude of the object in the Blender simulation was then determined through Equation \ref{eq:inst_mag}, which was simplified as the background sky signal ($S_{sky}$) is 0 in the Blender simulation.  Similar to the method applied when determining the apparent magnitude of RSOs in real imagery, this object can then be used as a reference point to determine the apparent magnitude of objects in Blender-generate imagery.\\

However, in the real world the stars have a fixed apparent magnitude for observers on earth, whilst the reference object that we have created using the flat facet model and Blender is only accurate for the specific distance of 10000\,m as the intensity of the light is subject to the inverse square law.  Thus, objects rendered in Blender at other distances will need to be corrected for this distance by dividing the extracted intensity by the square of the ratio of the actual distance used in Blender to the 10000\,m reference point.  A final correction must also be applied to account for the scale factor applied in Blender to the positions of the camera, sun and object, to reduce the distances required for computation in the Blender simulation.\\

These corrections can be applied to Equation \ref{eq:final_rel}, resulting in the final equation that was used to convert the sum of the Blender pixel intensities in an image to the apparent magnitude:
\begin{equation}
m_1 = m_{ff} - I_{m_b} + \left( -2.5\; \textrm{log}_{10} \left( \frac{B}{t_{exp}\cdot (\frac{d}{d_r}\cdot scale)^2} \right) \right)
\label{eq:final_equation}
\end{equation}
Where $m_{ff}$ is the apparent magnitude of the flat plate at 10000\,m determined using the flat facet model. $I_{m_b}$ is the Blender instrument magnitude of the flat plate ate 10000\,m. $B$ is the sum of the pixel intensity of the object extracted from the Blender render image, whilst $t_{exp}$ is the exposure time in seconds. $d$ is the distance between the object and the camera used in Blender, whilst $d_r$ is the distance used for the reference object which was 10000\,m.  Finally, $scale$ is the scale factor that was applied to the object, camera and sun ephemeris when inputting them into the Blender simulation, which is generally set to 1000.\\

Equation \ref{eq:final_equation} was used to convert the Blender pixel intensities extracted from the rendered images into a a calibrated apparent magnitude value that allowed direct comparison with real light curve data.  The extracted apparent magnitude for each image was recorded with the ephemeris information and parameters for the pass.  After processing, the Blender rendered images are deleted to reduce data storage requirements.  

\section{Extraction of Real Light Curves} \label{sec:Extraction}
\vspace{-0.2cm}
As part of the validation process for the light curve simulation environment an automated light curve extraction algorithm has been developed in collaboration with EOS and the Space Environment Research Centre (SERC).  EOS is an Australian based company who operate 6 optical telescopes at two sites in Australia and have provided the real data that is discussed in this paper. EOS telescopes primarily operate in a rate tracking state, resulting in the tracked object appearing circular in the image whilst background stars appear as streaks.\\

\begin{figure}
	\centering
	\begin{subfigure}{0.49\columnwidth}
		\includegraphics[width=\linewidth]{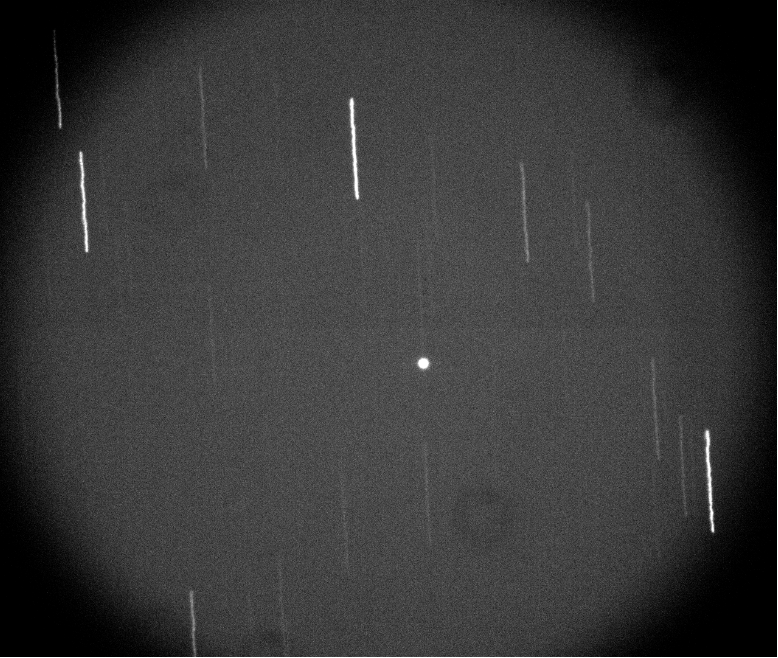}
		\caption{Telescope Imagery}
		\label{fig:eos_img}
	\end{subfigure}
	\begin{subfigure}{0.49\columnwidth}
		\includegraphics[width=\linewidth]{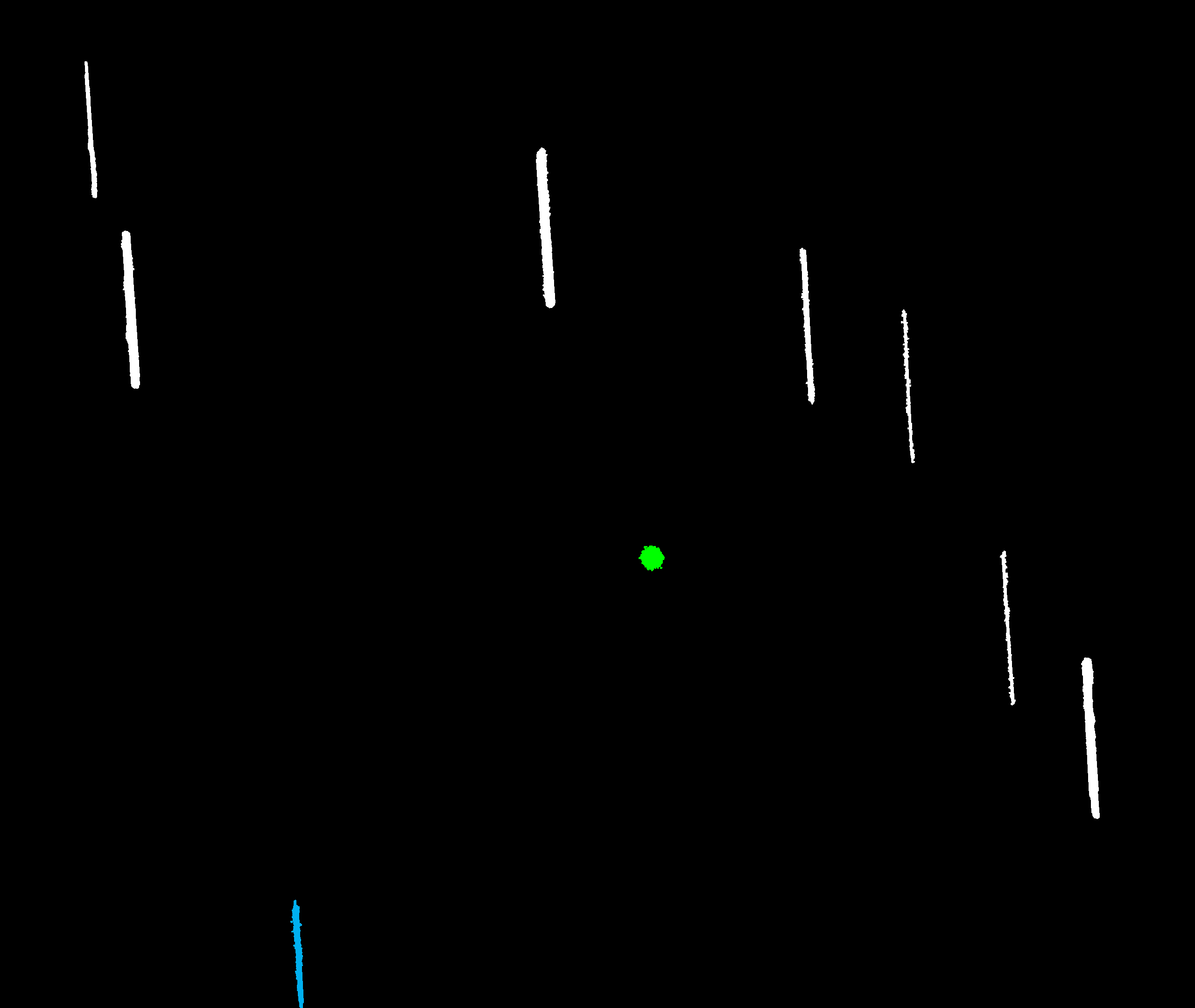}
		\caption{Object Detection Mask}
		\label{fig:mask}
	\end{subfigure}
	\caption{Non Resolved RSO}
	\label{fig:raw_EOS}
\end{figure}

Figure \ref{fig:eos_img} depicts a typical telescope image with an RSO near the centre of the image and a number of stars streaking through the image. The algorithm processes a sequence of such images and extracts the apparent magnitude for the RSO in each image which is then recorded to generate the light curve for the pass.\\

Detection of both RSOs and background stars are performed in each image using a Fast Fourier Transform (FFT) approach. An example of the mask created through this process is displayed in Figure \ref{fig:mask}.  In the mask the identified RSO is green, whilst complete streaking stars are white.  The blue streak in the bottom section of the image is identified as a partial star streak with only one endpoint.  The object is tracked through the sequence of images and its future position predicted, allowing for multiple RSOs within an image to be observed and tracked simultaneously.\\

The pixel counts recorded in the 16-bit depth grey-scale image captured by the telescope can be extracted for each object in the image using the aperture method \cite{romanishin_introduction_2006}.  These counts are then used to determine the instrument magnitude for both the RSOs and the stars in the image through the application of Equation \ref{eq:inst_mag}.  The apparent magnitude of the RSO is then determined by extracting the apparent magnitudes of the stars in the image from the UCAC4 catalogue and applying Equation \ref{eq:final_rel}.\\

\section{Results}
\vspace{-0.2cm}
\subsection{Extracted Real Light Curve}
\begin{figure*}
	\centering
	\includegraphics[width=1.0\textwidth]{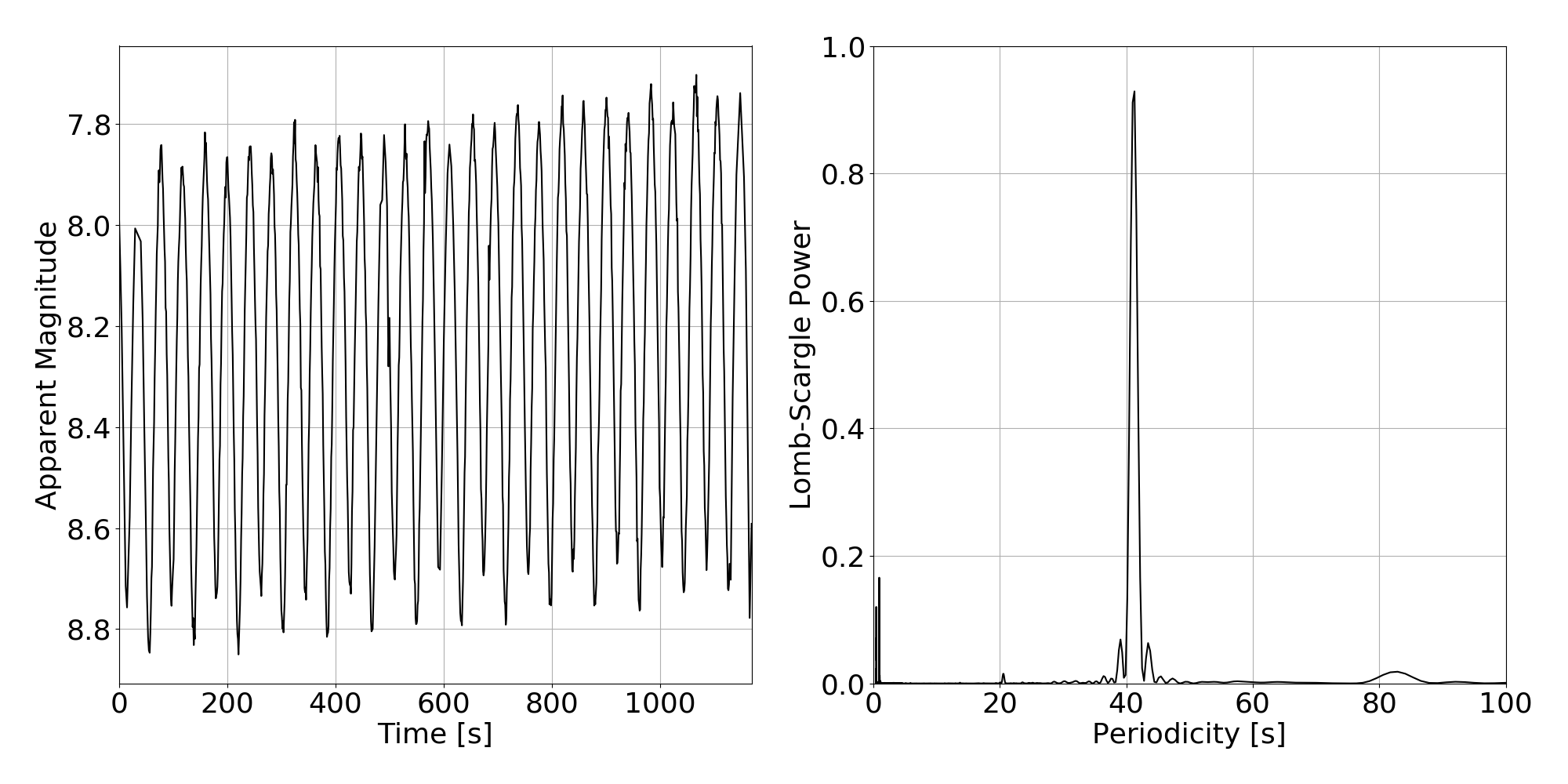}
	\caption{Extracted Light Curve and Spin Analysis}
	\label{fig:40108}
\end{figure*}

\vspace{-0.2cm}
The left hand side of Figure \ref{fig:40108} depicts a light curve that was extracted, using the algorithm outlined in Section \ref{sec:Extraction}, from observations of a SpaceX Falcon 9 stage 2 rocket body (NORAD ID 40108).  The imagery was captured at an average sampling rate of 1Hz for approximately 20 minutes on the 7th of July 2019 using the 0.7m telescope at Mt Stromlo.\\

The cyclical pattern present in the light curve indicates that object 40108 is rotating.  This was confirmed by applying the Lomb-Scargle analysis to the light curve, with a strong peak evident in the Lomb-Scargle power plot on the right hand side of Figure \ref{fig:40108} at a period of 41.3 seconds.  This period actually represents a half rotation of the rocket body, which, given its symmetrical nature, experiences bright and dim peaks twice a cycle and thus has a spin period of 82.6 seconds. This can also be observed through the pattern in the dips in the light curve visible in the lower section of the graph (higher magnitude) where every second dip is slightly dimmer. The consistent difference between these lower peaks indicates that the object is not symmetrical about all axis and increases confidence in the extracted spin period of 82.6 seconds.\\ 

A further trend visible in the light curve plot is the general increase in magnitude throughout the duration of the pass.  It is thought that this trend is a result of the gradual decrease in the phase angle measured during the pass. However, it could also be a function of a slow precession of the rocket body around a secondary axis.  The phase angle is the angle measured between the direction of the incoming flux (light from the sun) and the direction of the observer.\\

\subsection{Comparison between Simulated and Real Data}
\begin{figure*}
	\centering
	\begin{subfigure}{0.49\linewidth}
		\includegraphics[width=\linewidth]{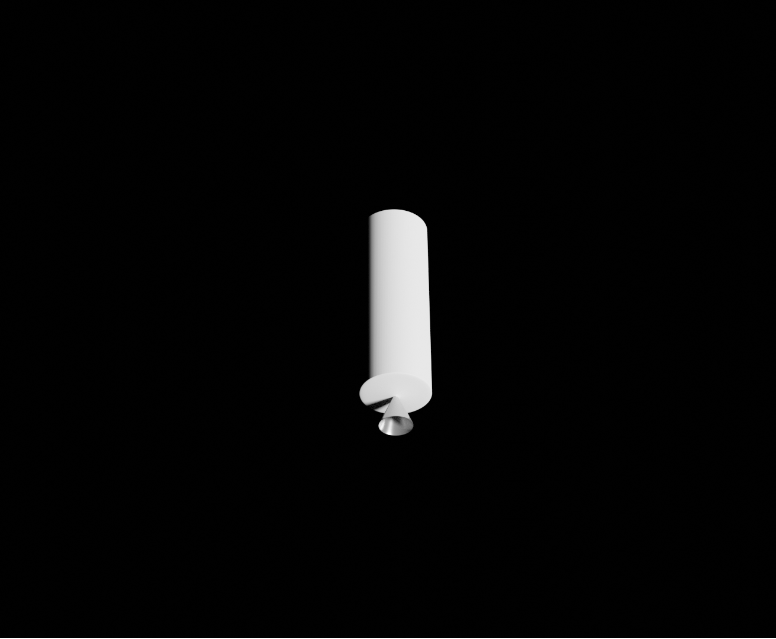}
		\caption{SpaceX Stage Falcon 2 Model}
		\label{fig:rb_snip}
	\end{subfigure}
	\begin{subfigure}{0.48\linewidth}
		\includegraphics[width=\linewidth]{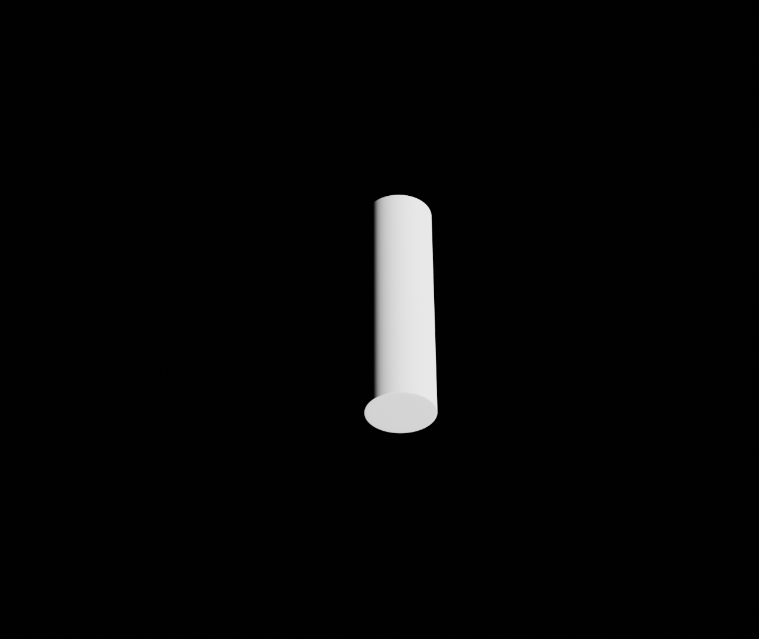}
		\caption{Simple Cylinder Model}
		\label{fig:cyl_snip}
	\end{subfigure}\\[1ex]
	\begin{subfigure}{\linewidth}
		\includegraphics[width=\linewidth]{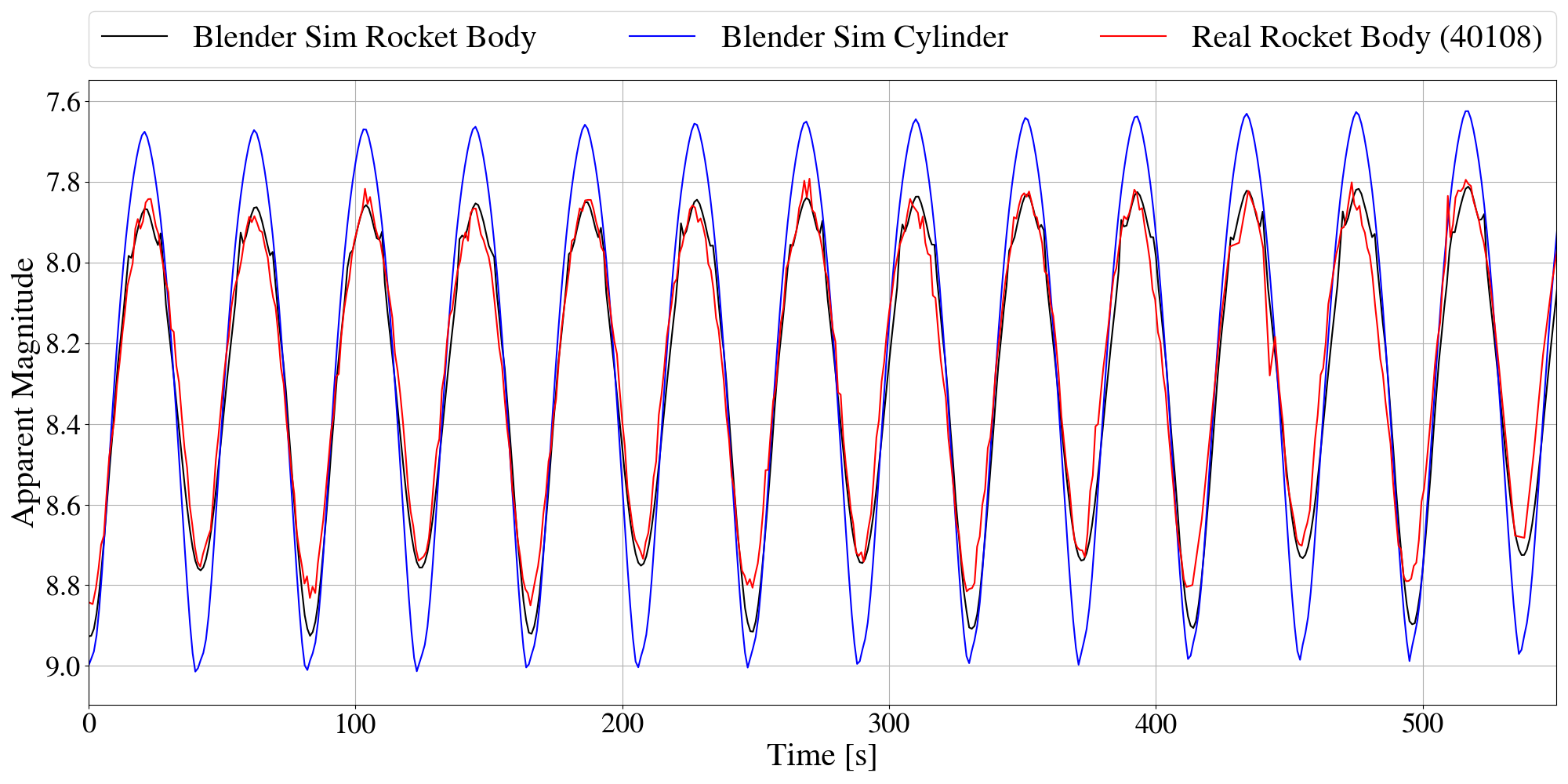}
		\caption{Comparison between Blender Simulated Data and Real Light Curve}
		\label{fig:simulation_comparison}
	\end{subfigure}	
	\caption{Initial Validation of Simulation Environment}
	\label{fig:validation}
\end{figure*}

\vspace{-0.2cm}
To compare the simulation environment with the extracted real light curve presented in Figure \ref{fig:40108}, a simple rocket body was modelled in Blender based on approximate dimensions of the second stage of the SpaceX Falcon 9 rocket.  It was modelled as a cylinder with a nozzle extruding at the base and a small nose cone at the tip.  A rendered version of the model used for the simulation is visible in Figure \ref{fig:rb_snip}. Additionally, a simple cylinder without the nose cone and the nozzle, depicted in \ref{fig:cyl_snip}, was simulated to be the same diameter and length as the SpaceX rocket body as a source of comparison.\\ 

The TLE for object 40108 was propagated in Python using SGP4 to the start of the observation period and positions for the rocket body, the observing telescope at Mt Stromlo and the sun were recorded at 1 second intervals for the duration of the observed pass.  This information was then input into the Blender simulation environment and the SpaceX model was rotated on both its x and y axes separately using the spin period of 82.6 seconds determined by the Lomb-Scargle analysis presented in Figure \ref{fig:40108}.\\

A good convergence was found with the real light curve for when the simulated SpaceX model was rotated about the x axis.  This light curve is presented in Figure \ref{fig:simulation_comparison} along with the extracted light curve and the simulated light curve from the cylinder rotated around its x axis. In particular, note that the pattern identified earlier where every second dip is dimmer is present in the simulated light curve for the rocket body.  Based on the simulation these dips were found to correspond with the nozzle of the rocket facing the telescope, whilst the slightly brighter dips (first evident at an approximate apparent magnitude of 8.75 at 40 seconds) occur when the nose cone of the rocket is facing the observer.  The peaks on the graph correspond to when the full face of the cylindrical part of the rocket body was observable to the telescope.\\

Whilst the light curve for the simulated simple cylinder has a similar oscillation to the other light curves, the pattern present in every second dip is not evident.  The cylinder has a higher apparent magnitude at the peaks as the length of the cylinder used was the total length of the rocket body including the nozzle and nose cone, so the face of the cylinder has a larger area than the face of the rocket body. The nozzle and nose cone for the rocket body were also found to be more reflective than the ends of the cylinder when facing the camera resulting in lower dips in the light curve of the cylinder.  Additionally, the peaks in both the rocket bodies light curves appear to have small features and disturbances whilst the cylinder has smooth peaks. This is hypothesised to be as a result of side-on reflections from the nozzle which are present for the rocket body but absent from the cylinder simulation.\\

Furthermore, all three light curves also feature the general trend of increased brightness that was identified previously.  This indicates that this trend is more likely to be a result of the decreasing phase angle than a complex spin effect, as both the simulated objects are only rotated around their x-axis.\\

\subsection{Motion Blur}
\begin{figure*}
	\centering
	\includegraphics[width=\textwidth]{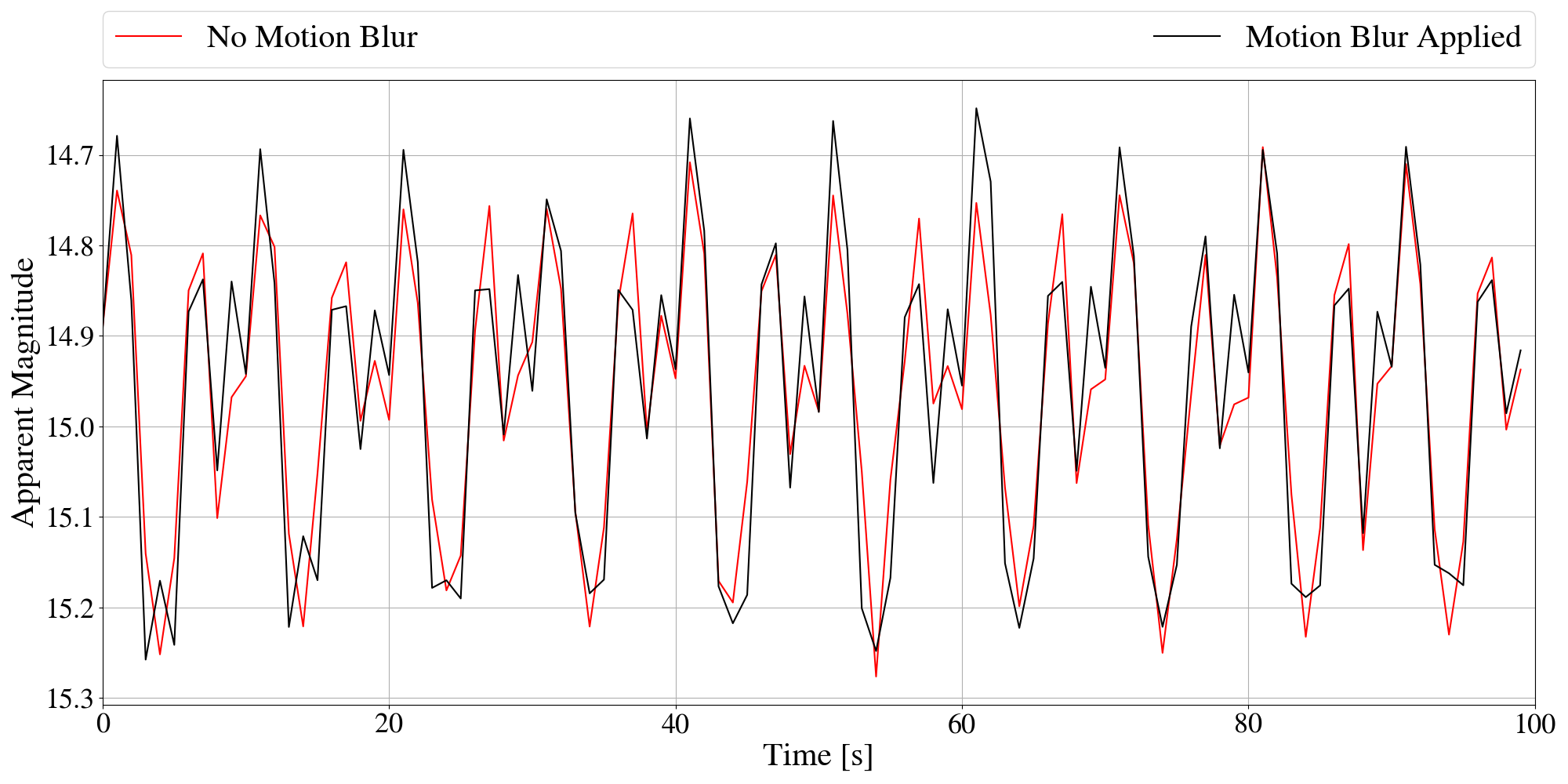}
	\caption{Light Curve Comparison of Motion Blur Implementation (3U CubeSat)}
	\label{fig:blur_lightcurve}
\end{figure*}

\vspace{-0.2cm}
Figure \ref{fig:blur_lightcurve} depicts a comparison for light curves generated using the Blender simulation environment for the case when motion blur is applied compared with no motion blur.  The light curves shown are extracted from a simulation of the 3U CubeSat depicted in Figure \ref{fig:blur_image}, rotating around its x axis with  spin period of 10 seconds.\\ 

While both light curves are quite similar in shape small features are different between them.  This is particularly evident at approximately 10 seconds into the pass where a distinct black peak is visible in the motion blur applied light curve that is not captured by the light curve without motion blur.  This peak is clearly visible in both light curves at other points during the pass in both light curves indicating that it is not a spurious peak. Additionally, further differences are notable between the two light curves in the magnitudes of the brighter peaks.\\

\section{Discussion}
\vspace{-0.2cm}
Although additional work is required to validate the Blender simulation environment, initial comparisons with rocket body light curves show good similarity between the real extracted data and the simulated data.  This close similarity, particularly when contrasted with the differences to the simulated light curve of the cylinder, emphasises the importance of using more complex models to generate realistic light curves.  Although the general shape of the cylinder is the same as the rocket body, the addition of the nozzle and nose cone has a significant effect on the resultant light curve.\\  

It should be noted that the dips corresponding to the nozzle are slightly dimmer in the simulation than in the real light curve.  This potentially indicates that the size of the nozzle should be increased or a more reflective material should be used in the simulation in order to better represent the real object.  Future development of the simulation environment will include creating realistic models of other space debris and performing further validation with extracted light curves to ensure that realistic light curves are being generated for a variety of shapes and models.\\

The inclusion of self-shadowing in the Blender modelling will allow for the simulation of more realistic light curves for convex objects and objects with large extrusions (e.g. solar panels). For these kinds of objects significant portions of the satellite may be shadowed by the solar panel or the body of the satellite may prevent reflections from the solar panel from reaching the observer. Additionally, motion blur has been shown to have an effect on the generated light curve, particularly for objects that are rapidly rotating or when longer exposures are used. Whilst these may seem like small details, it is expected that these are the kind of features that will allow a machine learning algorithm to differentiate between light curves associated with different shapes and models.\\

The visualisation aspect of the GUI built into Blender has also proved useful in analysing how small changes in the model affect the light curve.  This was evident in the work with the rocket body where analysis of the simulation allowed the orientation of the rocket body to be determined when specific features in the light curve were generated.\\

In terms of current differences between the simulated light curves and extracted light curves, images are currently simulated with no gap in exposure between images.  This differentiates them from EOS sensors which have a slight gap between images primarily to allow for readout time.  Subsequently, the real light curves may be missing information compared with the simulated data, particularly for objects with a high spin rate. Furthermore, the EOS sensors currently use an adaptive algorithm to determine the exposure time to use during a pass depending on the brightness of the object and the mode of the telescope.  This means that the exposure time sometimes changes during the course of a pass.  Future analysis will be done to determine whether either of these differences have a significant effect on the simulated light curve and if so they will be replicated in the simulation environment.\\

Finally, comparative to the flat facet model, one of the detriments of the Blender simulation environment is that it is significantly more computationally expensive.  Tests on a CPU with 12 cores simulating a 30 minute pass resulted in the generation of approximately 1 light curve  per hour, while a NVIDIA GeForce 1080 Ti graphics cards was able to generate a light curve in 15 minutes.  The simulation environment has now been modified to be run in parallel on the GPU, enabling an improvement to ~10 light curves per hour per GPU.\\

\section{Conclusion}
\vspace{-0.2cm}
A high-fidelity Blender based simulation environment has been developed for the generation of light curves of rotating space objects.  The simulator allows complex geometric models as input and includes self-shadowing as well as motion blur.  Initial validation through comparison with a real light curve of a SpaceX stage 2 rocket body show a close similarity between the simulated and real light curve.  Further validation will be performed in the future comparing simulated light curves with real light curves for other objects.\\  

Additionally, the simulator will be used to generate a database of realistic well-labelled light curves that can be used to train neural networks for automated space object characterisation.  Through the process of transfer learning these neural networks will then be applied to real data.  Increasing the similarity between the simulated and the real data reduces the requirement of extensive retraining during the transfer learning process when applying the network to real data.

%
%
%
%
%
%

\bibliographystyle{unsrturl}
\bibliography{IAC2019_PrePrint}

\begin{thebibliography}{10}

\bibitem{kessler_collision_1978}
Donald~J. Kessler and Burton~G. Cour‐Palais.
\newblock Collision frequency of artificial satellites: {The} creation of a
  debris belt.
\newblock {\em Journal of Geophysical Research: Space Physics},
  83(A6):2637--2646, June 1978.
\newblock URL:
  \url{http://agupubs.onlinelibrary.wiley.com/doi/abs/10.1029/JA083iA06p02637},
  \href {http://dx.doi.org/10.1029/JA083iA06p02637}
  {\path{doi:10.1029/JA083iA06p02637}}.

\bibitem{dianetti_observability_2018}
Andrew~D. Dianetti, Ryan Weisman, and John~L. Crassidis.
\newblock Observability {Analysis} for {Improved} {Space} {Object}
  {Characterization}.
\newblock {\em Journal of Guidance, Control, and Dynamics}, 41(1):137--148,
  January 2018.
\newblock URL: \url{https://arc.aiaa.org/doi/10.2514/1.G002229}, \href
  {http://dx.doi.org/10.2514/1.G002229} {\path{doi:10.2514/1.G002229}}.

\bibitem{silha_apparent_2018}
Jiri Silha, Jean-Noel Pittet, Michal Hamara, and Thomas Schildknecht.
\newblock Apparent rotation properties of space debris extracted from
  photometric measurements.
\newblock {\em Advances in Space Research}, 61(3):844--861, February 2018.
\newblock URL:
  \url{https://linkinghub.elsevier.com/retrieve/pii/S027311771730786X}, \href
  {http://dx.doi.org/10.1016/j.asr.2017.10.048}
  {\path{doi:10.1016/j.asr.2017.10.048}}.

\bibitem{linder_extraction_2015}
Esther Linder, Jiri Silha, Thomas Schildknecht, and Monika Hager.
\newblock Extraction of {Spin} {Periods} of {Space} {Debris} from {Optical}
  {Light} {Curves}.
\newblock 2015.

\bibitem{fruh_analysis_2010}
Carolin Fruh and Thomas Schildknecht.
\newblock Analysis of observed and simulated light curves of space debris.
\newblock In {\em Proceedings of 61st {International} {Astronautical}
  {Congress}}, volume~1, pages 194--203, Prague, Czech Republic, September
  2010.

\bibitem{jah_satellite_2007}
Moriba~K Jah and Ronald~A Madler.
\newblock Satellite {Characterization}: {Angles} and {Light} {Curve} {Data}
  {Fusion} for {Spacecraft} {State} and {Parameter} {Estimation}.
\newblock In {\em Proceedings of {AMOS} {Conference}}, pages 1--10, Maui,
  Hawaii, September 2007.

\bibitem{wetterer_attitude_2009}
Charles~J. Wetterer and Moriba~K. Jah.
\newblock Attitude {Determination} from {Light} {Curves}.
\newblock {\em Journal of Guidance, Control, and Dynamics}, 32(5):1648--1651,
  September 2009.
\newblock URL: \url{http://arc.aiaa.org/doi/10.2514/1.44254}, \href
  {http://dx.doi.org/10.2514/1.44254} {\path{doi:10.2514/1.44254}}.

\bibitem{linares_inactive_2012}
Richard Linares, Moriba~K Jah, and John~L Crassidis.
\newblock Inactive {Space} {Object} {Shape} {Estimation} {Via} {Astrometric}
  {And} {Photometric} {Data} {Fusion}.
\newblock {\em Advances in Astronautical Sciences}, 143:217--232, 2012.

\bibitem{linares_space_2014}
Richard Linares, Moriba~K. Jah, John~L. Crassidis, and Christopher~K.
  Nebelecky.
\newblock Space {Object} {Shape} {Characterization} and {Tracking} {Using}
  {Light} {Curve} and {Angles} {Data}.
\newblock {\em Journal of Guidance, Control, and Dynamics}, 37(1):13--25,
  January 2014.
\newblock URL: \url{http://arc.aiaa.org/doi/10.2514/1.62986}, \href
  {http://dx.doi.org/10.2514/1.62986} {\path{doi:10.2514/1.62986}}.

\bibitem{furfaro_space_2018}
Roberto Furfaro, Richard Linares, and Vishnu Reddy.
\newblock Space {Objects} {Classification} via {Light}-{Curve} {Measurements}:
  {Deep} {Convolutional} {Neural} {Networks} and {Model}-based {Transfer}
  {Learning}.
\newblock In {\em Proceedings of {AMOS} {Conference}}, pages 1--17, Maui,
  Hawaii, September 2018.

\bibitem{goodfellow_deep_2017}
Ian Goodfellow, Yoshua Bengio, and Aaron Courville.
\newblock {\em Deep {Learning}}.
\newblock Adaptive {Computation} and {Machine} {Learning} {Series}. MIT Press,
  Cambridge, MA, 1st edition, 2017.

\bibitem{chollet_deep_2018}
François Chollet.
\newblock {\em Deep learning with {Python}}.
\newblock Manning, Shelter Island, NY, 2018.
\newblock OCLC: 1019988472.

\bibitem{linares_space_2016}
Richard Linares and Roberto Furfaro.
\newblock Space {Object} {Classification} {Using} {Deep} {Convolutional}
  {Neural} {Networks}.
\newblock In {\em Information {Fusion} ({FUSION}), 2016 19th {International}
  {Conference} on}, pages 1140--1146. IEEE, August 2016.

\bibitem{jia_space_2018}
B.~Jia, K.~D. Pham, E.~Blasch, Z.~Wang, D.~Shen, and G.~Chen.
\newblock Space {Object} {Classification} using {Deep} {Neural} {Networks}.
\newblock In {\em 2018 {IEEE} {Aerospace} {Conference}}, pages 1--8, March
  2018.
\newblock \href {http://dx.doi.org/10.1109/AERO.2018.8396567}
  {\path{doi:10.1109/AERO.2018.8396567}}.

\bibitem{huo_classification_2019}
Yurong Huo, Zhi Li, Yuqiang Fang, and Feng Zhang.
\newblock Classification for geosynchronous satellites with deep learning and
  multiple kernel learning.
\newblock {\em Applied Optics}, 58(21):5830, July 2019.
\newblock URL:
  \url{https://www.osapublishing.org/abstract.cfm?URI=ao-58-21-5830}, \href
  {http://dx.doi.org/10.1364/AO.58.005830} {\path{doi:10.1364/AO.58.005830}}.

\bibitem{krizhevsky_imagenet_2017}
Alex Krizhevsky, Ilya Sutskever, and Geoffrey~E. Hinton.
\newblock {ImageNet} classification with deep convolutional neural networks.
\newblock {\em Communications of the ACM}, 60(6):84--90, May 2017.
\newblock URL: \url{http://dl.acm.org/citation.cfm?doid=3098997.3065386}, \href
  {http://dx.doi.org/10.1145/3065386} {\path{doi:10.1145/3065386}}.

\bibitem{mahajan_exploring_2018}
Dhruv Mahajan, Ross Girshick, Vignesh Ramanathan, Kaiming He, Manohar Paluri,
  Yixuan Li, Ashwin Bharambe, and Laurens van~der Maaten.
\newblock Exploring the {Limits} of {Weakly} {Supervised} {Pretraining}.
\newblock {\em arXiv:1805.00932 [cs]}, May 2018.
\newblock arXiv: 1805.00932.
\newblock URL: \url{http://arxiv.org/abs/1805.00932}.

\bibitem{furfaro_resident_2016}
Roberto Furfaro, Richard Linares, David Gaylor, Moriba Jah, and Ramona Walls.
\newblock Resident {Space} {Object} {Characterization} and {Behavior}
  {Understanding} via {Machine} {Learning} and {Ontology}-based {Bayesian}
  {Networks}.
\newblock In {\em Advanced {Maui} {Optical} and {Space} {Surveillance} {Tech}.
  {Conf}.({AMOS})}, Maui, Hawaii, September 2016.

\bibitem{ashikhmin_anisotropic_2000}
Michael Ashikhmin and Peter Shirley.
\newblock An {Anisotropic} {Phong} {Light} {Reflection} {Model}.
\newblock Technical report, Journal of Graphics Tools, 2000.

\bibitem{hubaux_symplectic_2012}
Ch. Hubaux, A.~Lemaître, N.~Delsate, and T.~Carletti.
\newblock Symplectic integration of space debris motion considering several
  {Earth}’s shadowing models.
\newblock {\em Advances in Space Research}, 49(10):1472--1486, May 2012.
\newblock URL:
  \url{https://linkinghub.elsevier.com/retrieve/pii/S0273117712001081}, \href
  {http://dx.doi.org/10.1016/j.asr.2012.02.009}
  {\path{doi:10.1016/j.asr.2012.02.009}}.

\bibitem{kanzlar_space_2015}
R.~Kanzlar, J.~Silha, T~Schildknecht, B.~Fritsche, T.~Lips, and H.~Krag.
\newblock Space {Debris} {Attitude} {Simulation} - {iOTA} ({In}-{Orbit}
  {Tumbling} {Analysis}).
\newblock In {\em Advanced {Maui} {Optical} and {Space} {Surveillance} {Tech}.
  {Conf}.({AMOS})}, Hawaii, 2015.

\bibitem{blender}
Blender~Online Community.
\newblock {\em Blender - a 3D modelling and rendering package}.
\newblock Blender Foundation, Blender Institute, Amsterdam, 2019.
\newblock URL: \url{http://www.blender.org}.

\bibitem{mehta_photometric_2018}
Piyush~M. Mehta, Richard Linares, and Andrew~C. Walker.
\newblock Photometric {Data} from {Nonresolved} {Objects} for {Improved} {Drag}
  and {Reentry} {Prediction}.
\newblock {\em Journal of Spacecraft and Rockets}, 55(4):959--970, July 2018.
\newblock URL: \url{https://arc.aiaa.org/doi/10.2514/1.A33825}, \href
  {http://dx.doi.org/10.2514/1.A33825} {\path{doi:10.2514/1.A33825}}.

\bibitem{nasa_models}
Kristen Erikson.
\newblock {\em NASA 3D Models}.
\newblock National Aeronautics and Space Admistration.
\newblock URL: \url{https://nasa3d.arc.nasa.gov/models/sort_name}.

\bibitem{romanishin_introduction_2006}
W.~Romanishin.
\newblock {\em An {Introduction} to {Astronomical} {Photometry} {Using}
  {CCDs}}.
\newblock University of Oklahoma, October 2006.

\end{thebibliography}

\end{singlespace}
\end{document}